\documentclass{aa}    

\usepackage{graphicx,url,twoopt,natbib}
\usepackage[varg]{txfonts}           
\usepackage[usenames]{color}
\usepackage{acronym}      
\usepackage[titles]{tocloft}
\usepackage{stdclsdv} 
\usepackage{hyperref} 
\hypersetup{
  colorlinks=true,  
  urlcolor=rrblue,    
  citecolor=rrblue,   
  linkcolor=rrblue    
}

\definecolor{bllsht}{rgb}{0.40, 0.05, 0.01} 
\long\def\TODO#1TODO{\par\vspace{0.5ex}\noindent
  {\Large \textcolor{red}{@~}}\color{bllsht}{#1}\color{black}\\[0.5ex]}

\definecolor{rrblue}{rgb}{0.15,0.0,0.8} 

\makeatletter
\newcommand{\bibnote}[2]{\global\@namedef{#1note}{#2~}}
\newcommand{\biblink}[2]{\global\@namedef{#1link}{#2}}
\makeatother

\definecolor{amber}{rgb}{1.0, 0.49, 0.0}

\bibpunct{(}{)}{;}{a}{}{,}    
\makeatletter
  \protected\def\sppresslink{\def\hyper@linkstart##1##2{}\let\hyper@linkend\@empty}
  \newcommandtwoopt{\citeads}[3][][]{%
   \href{http://ui.adsabs.harvard.edu/abs/#3/abstract}%
        {\sppresslink \citealp[#1][#2]{#3}}
   \biblink{#3}{\href{http://ui.adsabs.harvard.edu/abs/#3/abstract}{ADS}}}
 \newcommandtwoopt{\citepads}[3][][]{%
   \href{http://ui.adsabs.harvard.edu/abs/#3/abstract}%
        {\sppresslink \citep[#1][#2]{#3}}
   \biblink{#3}{\href{http://ui.adsabs.harvard.edu/abs/#3/abstract}{ADS}}}
 \newcommandtwoopt{\citetads}[3][][]{%
   \href{http://ui.adsabs.harvard.edu/abs/#3/abstract}%
        {\sppresslink \citet[#1][#2]{#3}}
  \biblink{#3}{\href{http://ui.adsabs.harvard.edu/abs/#3/abstract}{ADS}}}
 \newcommandtwoopt{\citeyearads}[3][][]{%
   \href{http://ui.adsabs.harvard.edu/abs/#3/abstract}%
        {\sppresslink \citeyear[#1][#2]{#3}}
   \biblink{#3}{\href{http://ui.adsabs.harvard.edu/abs/#3/abstract}{ADS}}}
\makeatother

\newcount\longrefs
\def\adv{\ifnum\longrefs=1 {Adv.\ Space Res.} \else 
                           {Adv.\ Sp'\ Res.}\fi}
\def\aap{\ifnum\longrefs=1 {Astron.\ Astrophys.}\else 
                           {A\hbox{\rm \&}A}\fi}
\def\aapr{\ifnum\longrefs=1 {Astron.\ Astrophys.\ Rev.}\else 
                            {A\hbox{\rm \&}AR}\fi}
\def\aaps{\ifnum\longrefs=1 {Astron.\ Astrophys.\ Suppl.}\else 
                            {A\hbox{\rm \&}A Suppl.}\fi}
\def\actaa{\ifnum\longrefs=1 {Acta Astronomica}\else
                            {Acta Astron.}\fi}
\def\aipcs{\ifnum\longrefs=1 {Am.\ Inst.\ Phys.\ Conf.\ Series}\else
                             {AIP Conf.\ Ser.}\fi}
\def\aj{\ifnum\longrefs=1 {Astron.\ J.}\else 
                          {AJ}\fi} 
\def\ao{\ifnum\longrefs=1 {Applied Optics}\else 
                           {Appl.\ Opt.}\fi} 
\def\aspcs{\ifnum\longrefs=1 {Astron.\ Soc.\ Pacific Conf.\ Series}\else 
                           {ASP Conf.\ Ser.}\fi} 
\def\apj{\ifnum\longrefs=1 {Astrophys.\ J.}\else 
                           {ApJ}\fi} 
\def\apjl{\ifnum\longrefs=1 {Astrophys.\ J. Lett.}\else 
                            {ApJL}\fi} 
\def\aplett{\ifnum\longrefs=1 {Astrophys.\ J. Lett.}\else 
                            {ApJ}\fi} 
\def\apjs{\ifnum\longrefs=1 {Astrophys.\ J. Suppl.}\else 
                            {ApJS}\fi}
\def\apss{\ifnum\longrefs=1 {Astrophys.\ Space Sci.}\else 
                            {Astrophys.\ Space Sci.}\fi}
\def\araa{\ifnum\longrefs=1 {Ann.\ Rev.\ Astron.\ Astrophys.}\else 
                            {ARA\hbox{\rm \&}A}\fi}
\def\azh{\ifnum\longrefs=1 {Astronomicheskii Zhurnal}\else 
                            {Astron.\ Zhur.}\fi}
\def\baas{\ifnum\longrefs=1 {Bull.\ Am.\ Astron.\ Soc.}\else 
                            {BAAS}\fi}
\def\bain{\ifnum\longrefs=1 {Bull.\ Astronom.\ Institutes Netherlands}\else
                            {Bull.\ Astr.\ Inst.\ Neth.}\fi}
\def\cjaa{\ifnum\longrefs=1 {Chin.\ J.\ Astron.\ Astrophys.}\else 
                            {Chin.\ J.\ A\&A}\fi}
\def\gca{\ifnum\longrefs=1 {Geochim.\ Cosmochim.\ Acta}\else 
                           {Geochim.\ Cosmochim.\ Acta}\fi}
\def\grl{\ifnum\longrefs=1 {Geophys.\ Res.\ Lett.}\else 
                           {Geoph.\ Res.\ Lett.}\fi}
\def\iaucirc{\ifnum\longrefs=1 {IAU Circulars}\else 
                          {IAU Circ.}\fi}
\def\icarus{\ifnum\longrefs=1 {Icarus}\else 
                          {Icarus}\fi}
\def\ip{\ifnum\longrefs=1 {in press}\else 
                          {in press}\fi}
\def\jcap{\ifnum\longrefs=1 {Jour.\ Cosmology Astropart.\ Phys.}\else 
                          {JCAP}\fi}
\def\jgr{\ifnum\longrefs=1 {J.\ Geophys.\ Res.}\else 
                           {J.\ Geophys.\ Res.}\fi}  
\def\jrasc{\ifnum\longrefs=1 {J.\ Royal Astron.\ Soc.\ Canada}\else 
                             {JRAS Can.}\fi}  
\def\memsai{\ifnum\longrefs=1 {Mem.~Soc.~Astron.~Italiana}\else
                              {MmSAI}\fi}
\def\mnras{\ifnum\longrefs=1 {Mon.\ Not.\ Roy.\ Astron.\ Soc.}\else 
                             {MNRAS}\fi} 
\def\na{\ifnum\longrefs=1 {New Astronomy}\else 
                          {New Astron.}\fi}
\def\nar{\ifnum\longrefs=1 {New Astronomy rev.}\else 
                           {New Astron.\ Rev.}\fi}
\def\nat{\ifnum\longrefs=1 {Nature}\else 
                           {Nat}\fi}
\def\pasa{\ifnum\longrefs=1 {Pub.\ Astron.\ Soc.\ Australia}\else 
                            {PASA}\fi} 
\def\pasj{\ifnum\longrefs=1 {Pub.\ Astron.\ Soc.\ Japan}\else 
                            {PASJ}\fi} 
\def\pasp{\ifnum\longrefs=1 {Pub.\ Astron.\ Soc.\ Pacific}\else 
                            {PASP}\fi} 
\def\physscr{\ifnum\longrefs=1 {Physica Scripta}\else 
                               {Phys.\ Scrip.}\fi} 
\def\planss{\ifnum\longrefs=1 {Planetary \& Space Science}\else 
                              {Plan. \& Space Sci.}\fi} 
\def\pre{\ifnum\longrefs=1 {Phys.\ Rev.\ E}\else
                           {Phys.\ Rev.\ E}\fi}
\def\procspie{\ifnum\longrefs=1 {Proc.\ SPIE}\else 
                                {Proc.\ SPIE}\fi} 
\def\qjras{\ifnum\longrefs=1 {Quarterly J.\ Royal Astron.\ Soc.}\else 
                             {QJRAS}\fi} 
\def\rmxaa{\ifnum\longrefs=1 {Revista Mexicana de Astron.\ y Astrofys.}\else 
                             {RMxAA}\fi} 
\def\sa{\ifnum\longrefs=1 {Soviet Astron..}\else 
                          {Sov.\ Astron.}\fi}
\def\skytel{\ifnum\longrefs=1 {Sky \& Telescope}\else 
                              {Sky \& Tel.}\fi} 
\def\solphys{\ifnum\longrefs=1 {Solar Phys.}\else 
                               {SoPh}\fi}
\def\sovast{\ifnum\longrefs=1 {Soviet Astron.}\else 
                              {Sov.\ Ast.}\fi}
\def\ssr{\ifnum\longrefs=1 {Space Sci.\ Rev.}\else 
                           {Space Sci.\ Rev.}\fi}
\def\zap{\ifnum\longrefs=1 {Zeitschr.\ f.\ Astrophysik}\else
                               {Z.\ Astrophys.}\fi}

\newacro{AA}{Astronomy \& Astrophysics}  
\newacro{ADS}{Astrophysics Data System}
\newacro{AIA}{Atmospheric Imaging Assembly}
\newacro{ALMA}{Atacama Large Millimeter/submillimeter Array}
\newacro{AO}{adaptive optics}
\newacro{ApJ}{Astrophysical Journal}
\newacro{AR}{active region}
\newacro{bb}{bound-bound}
\newacro{bf}{bound-free}
\newacro{BFI}{Broad-band Filter Imager}
\newacro{CE}{coronal equilibrium}
\newacro{CfA}{Center for Astrophysics}
\newacro{CME}{coronal mass ejection}
\newacro{CRD}{complete redistribution}
\newacro{CRISP}{CRisp Imaging SpectroPolarimeter}
\newacro{CRISPEX}{CRisp SPectral EXplorer}
\newacro{CS}{coherent scattering}
\newacro{DEM}{Differential Emission Measure}
\newacro{DF}{dynamic fibril}
\newacro{DKIST}{Daniel K. Inouye Solar Telescope}
\newacro{DLR}{Deutsches Zentrum f\"ur Luft- und Raumfahrt}
\newacro{DOT}{Dutch Open Telescope}
\newacro{DST}{Richard B. Dunn Solar Telescope}   
\newacro{EB}{Ellerman bomb}
\newacro{EDP}{\'{E}dition Diffusion Presse}  
\newacro{EIT}{Extreme ultraviolet Imaging Telescope}
\newacro{EPIC}{European participation in Solar-C}
\newacro{ERC}{European Research Council}
\newacro{ESA}{European Space Agency}
\newacro{EST}{European Solar Telescope}
\newacro{EUV}{extreme ultraviolet}
\newacro{FAF}{flaring active-region fibril}
\newacro{ff}{free-free}
\newacro{FITS}{Flexible Image Transport System}
\newacro{FOV}{field of view}
\newacro{fov}{field of view}
\newacro{FWHM}{full width at half maximum}
\newacro{HAO}{High Altitude Observatory}
\newacro{HD}{hydrodynamics}
\newacro{Hi-C}{High Resolution Coronal Imager Sounding Rocket}
\newacro{HMI}{Helioseismic and Magnetic Imager}
\newacro{IAA}{Instituto de Astrof\'{i}sica de Andaluc\'{i}a}
\newacro{IAC}{Instituto de Astrof\'{i}sica de Canarias}
\newacro{IAS}{Institut d'Astrophysique Spatiale}
\newacro{IAU}{International Astronomical Union}
\newacro{IBIS}{Interferometric Bi-dimensional Spectrometer}
\newacro{IDL}{Interactive Data Language}
\newacro{IMaX}{Imaging Magnetograph eXperiment}
\newacro{INAF}{Istituto Nazionale di Astrofisica}
\newacro{IB}{IRIS bomb}
\newacro{IR}{infrared}
\newacro{IRIS}{Interface Region Imaging Spectrograph}
\newacro{ISAS}{Institute of Space and Astronautical Science}
\newacro{ISP}{Institute for Solar Physics}
\newacro{ISS}{International Space Station}
\newacro{ISSI}{International Space Science Institute}
\newacro{ITA}{Institute for Theoretical Astrophysics}
\newacro{JAXA}{Japan Aerospace Exploration Agency}
\newacro{JSOC}{Joint Science Operations Center}
\newacro{KIS}{Kiepenheuer--Institut f\"{u}r Sonnenphysik}
\newacro{KPNO}{Kitt Peak National Observatory}
\newacro{LASP}{Laboratory for Atmospheric and Space Physics}
\newacro{LC}{liquid cristal}
\newacro{LMSAL}{Lockheed Martin Solar and Astrophysics Labratory}
\newacro{LOS}{line of sight}
\newacro{LTE}{local thermodynamic equilibrium}
\newacro{MC}{magnetic concentration}
\newacro{MCAO}{multi-conjugate adaptive optics} 
\newacro{MDI}{Michelson Doppler Imager}
\newacro{ME}{Milne-Eddington} 
\newacro{MHD}{magnetohydrodynamics}
\newacro{MOMFBD}{Multi-Object Multi-Frame Blind Deconvolution}
\newacro{MPE}{Max--Planck--Institut f\"ur extraterrestrische Physik}
\newacro{MPG}{Max--Planck--Gesellschaft}
\newacro{MPS}{Max Planck Institute for Solar System Research}
\newacro{MSSL}{Mullard Space Science Laboratory}
\newacro{MTF}{modulation transfer function}
\newacro{NAOJ}{National Astronomical Observatory of Japan}
\newacro{NASA}{National Aeronautics and Space Administration}
\newacro{NIST}{National Institute of Standards and Technology}
\newacro{NLTE}{non-local thermodynamic equilibrium}
\newacro{NLFFF}{non-linear force-free field}
\newacro{NOAA}{National Oceanic and Atmospheric Administration}
\newacro{non-E}{non-equilibrium}
\newacro{NSO}{National Solar Observatory}
\newacro{NWO}{Netherlands Organisation for Scientific Research}
\newacro{PHE}{propagating heating event}
\newacro{PRD}{partial redistribution}
\newacro{PROBA2}{PRoject for OnBoard Autonomy}
\newacro{PSBE}{post Saha-Boltzmann extinction}
\newacro{PSF}{point spread function}
\newacro{QS}{quiet Sun}
\newacro{QSEB}{quiet-Sun Ellerman-like brightening} 
\newacro{RAL}{Rutherford Appleton Laboratory}
\newacro{RBE}{rapid blue-shifted excursion}
\newacro{R-MHD}{radiation hydrodynamics}
\newacro{rms}{root mean square}
\newacro{RMS}{root mean square}
\newacro{ROB}{Royal Observatory of Belgium}
\newacro{ROI}{region of interest}
\newacro{RRE}{rapid red-shifted excursion}
\newacro{RTE}{radiative transfer equation}
\newacro{RTSA}{Radiative Transfer in Stellar Atmospheres}
\newacro{SCF}{slender \CaIIH\ fibril}
\newacro{SE}{statistical equilibrium}
\newacro{SB}{Saha Boltzmann}
\newacro{SDO}{Solar Dynamics Observatory}
\newacro{SJI}{slit-jaw image}
\newacro{SLI}{slit image}
\newacro{SNR}{signal-to-noise ratio}
\newacro{SO}{Solar Orbiter}
\newacro{SoHO}{Solar and Heliospheric Observatory}
\newacro{SP}{Spectropolarimeter}
\newacro{SST}{Swedish 1-m Solar Telescope}
\newacro{SUMER}{Solar Ultraviolet Measurements of Emitted Radiation}
\newacro{SUFI}{Sunrise Filter Imager}
\newacro{SVD}{singular value decomposition}
\newacro{SVST}{Swedish Vacuum Solar Telescope}
\newacro{STX}{Solar Telescope X}
\newacro{THEMIS}{T\'{e}lescope H\'{e}liographique pour l'Etude du 
   Magn\'{e}tisme et des Instabilit\'{e} Solaires}     
\newacro{TR}{transition region}
\newacro{TRACE}{Transition Region and Coronal Explorer}
\newacro{TSI}{total solar irradiance}
\newacro{UT}{Universal Time}
\newacro{UV}{ultraviolet}
\newacro{VAULT}{Very high angular resolution ultraviolet telescope}
\newacro{VIRGO}{Variability of solar IRradiance and Gravity Oscillations}
\newacro{VTT}{Vacuum Tower Telescope}    
\newacro{XRT}{X-Ray Telescope}

\long\def\startignore #1\stopignore{}   

\def\specchar#1{\uppercase{#1}}    
\def\specand{ and }                
\def\specand{\,\&\,}               

\def\CaII{\mbox{Ca\,\specchar{ii}}}

\def\FeI{\mbox{Fe\,\specchar{i}}}

\def\FeIX{\mbox{Fe\,\specchar{ix}}}
\def\FeX{\mbox{Fe\,\specchar{x}}}

\def\HeII{\mbox{He\,\specchar{ii}}}

\def\MgI{\mbox{Mg\,\specchar{i}}}

\def\CaIIH{\mbox{Ca\,\specchar{ii}\,\,H}}
\def\CaIIHK{\mbox{Ca\,\specchar{ii}\,\,H{\specand}K}}

\def\level #1 #2#3#4{$#1 \; ^{#2} \mbox{#3} ^{#4}$}   

\spacing{1.01}   

\def\paragraphrr#1{\paragraph*{#1.~~} \addcontentsline{toc}{subsection}{#1}}

\begin{document}  

\twocolumn[{%
{\em \Large Review in ``Solar Magnetic Variability and
Climate''\\[0.5ex]
C. de Jager, S.~Duhau, A.C.T.~Nieuwenhuizen, 2020}\\[1.3ex]
{\large \sf
\href{https://robrutten.nl/rrweb/rjr-pubstuff/bookcdej/solar-variability-and-cimate-content.pdf}{book
contents} ~~~~~~~~~\href{https://stipmedia.nl}{Stip Media}
~~~~~~~~~order: vincent@stipmedia.nl}

\begin{center}
\vspace*{4ex} {\LARGE \bf  Small-scale solar surface magnetism}\\[2ex]
{\large \bf Robert J. Rutten$^{1, 2, 3}$}\\[2ex]
\begin{minipage}[t]{15cm} \small 
\mbox{}\hspace{20ex}$^1$ Lingezicht Astrophysics,
Deil, The Netherlands\\
\mbox{}\hspace{20ex}$^2$ Institute of Theoretical Astrophysics,
University in Oslo, Oslo, Norway\\
\mbox{}\hspace{20ex}$^3$ Rosseland Centre for Solar Physics,
University in Oslo, Oslo, Norway\\

{\bf Abstract.~} \small This contribution to ``Solar Magnetic
Variability and Climate'' reviews small-scale magnetic features on the
solar surface, in particular the strong-field but tiny magnetic
concentrations that constitute network and plage and represent most
magnetism outside sunspots and filaments. 
Where these are mostly of the same polarity, as in active-region
plage, their occurrence varies with the activity variations measured
by the sunspot number, but when they appear bipolar-mixed on small
scales they can also result from granular-scale dynamo action that
does not vary with the cycle.
\end{minipage}
\end{center}
}] 

\noindent
\begin{minipage}[t]{0.98\columnwidth}  
  {\small \tableofcontents}
  \label{sec:contents}
\end{minipage}

\begin{figure*}[hbtp]
  \centering
  \includegraphics[width=\textwidth]{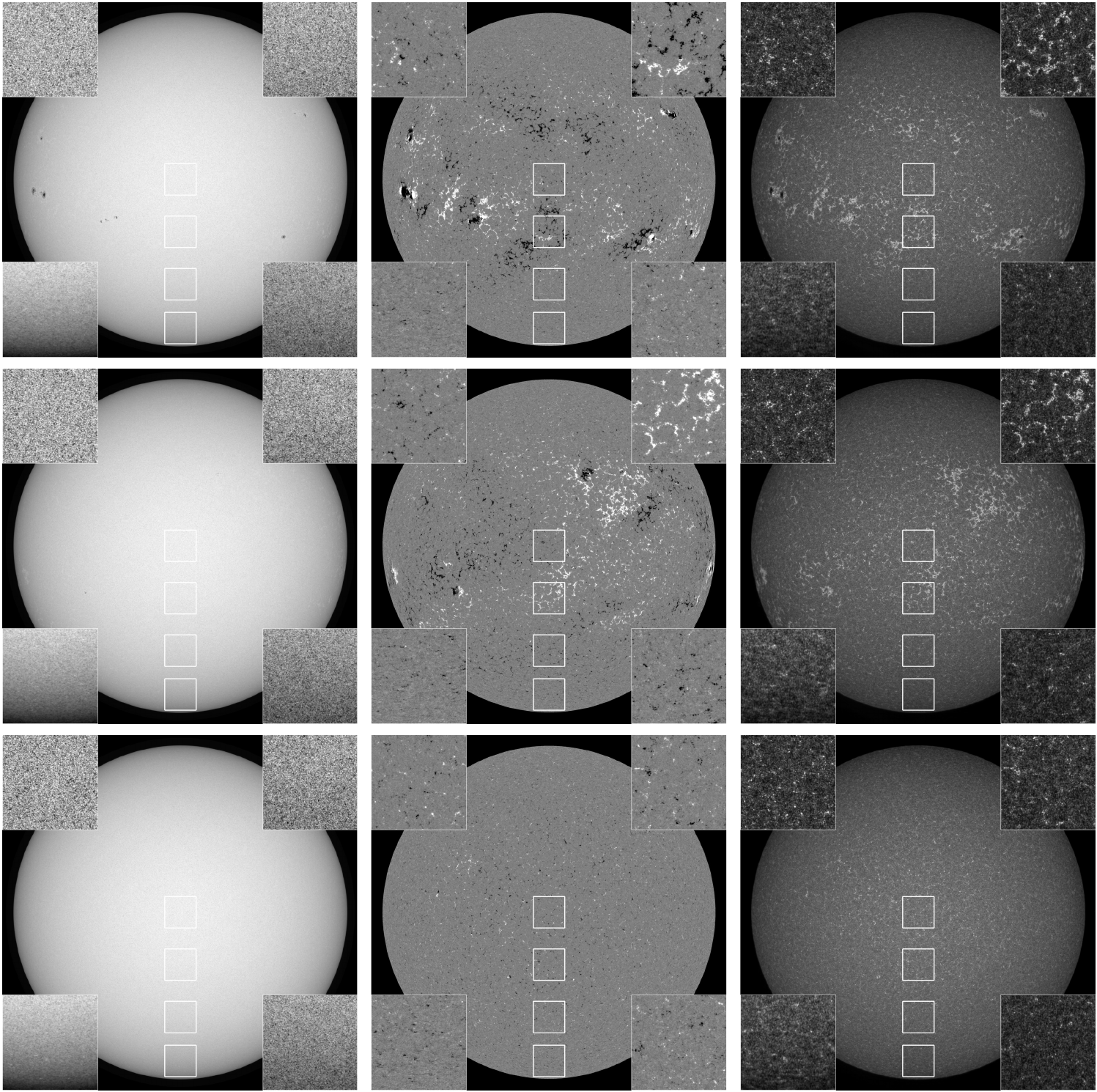}
  \caption[]{\label{fig:rrfig1} %
  Overview of characteristic solar surface magnetism patterns using
  full-disk images from the Solar Dynamics Observatory. 
  The rows are for three dates: April 29 2014 with much activity
  midway Cycle 24, April 29 2015 at declining activity with only one
  tiny sunspot, and April 29 2019 at low activity. 
  Each row contains the HMI continuum image, the HMI line-of-sight
  magnetogram and the AIA 1700\,\AA\ image collected at 03:00~UT. 
  The white boxes specify cut-outs that are magnified in the corners,
  in clockwise order for the four boxes from disk center to the South
  limb. 
  The HMI magnetograms are clipped at 100~Gauss apparent field
  strength to enhance network and plage fields. 
  The cut-outs are each greyscaled to maximum contrast.
  }
\end{figure*}

\paragraphrr{Magnetic concentrations constituting network, plage, faculae}

Solar surface magnetism exhibits a major discrete component in
addition to sunspots in active regions. 
It consists of small magnetic concentrations (henceforth MC) with
kilogauss field strength, spread over the Sun in irregular patterns. 
They represent a low but long tail in the size hierarchy of
strong-field features that emanate more or less upright (radially,
locally vertically) through the surface
(\citeads{1967SoPh....1..478Z}). 
For reviews see \citetads{2009SSRv..144..275D} 
and
\citetads{2017SSRv..210..275B}.  

Sunspot umbrae are the largest vertical flux concentrations. 
In their surrounding penumbrae the fields partly fan out to larger
height and partly bend back to below the surface. 
Next come magnetic pores: mini-umbrae without penumbra, of granular or
slightly larger size, also appearing dark on the surface. 
The yet smaller MCs also represent upright field bundles but with much
smaller cross-section at the surface, at most a few hundred km
diameter, usually less. 
Hence, telescopes need sub-arcsecond or better angular resolution to
observe them (1 arcsec measures about 700 km on the Sun). 
Even then their fine structure is often not resolved. 
They therefore became known as magnetic ``elements'' or ``points''
(magnetic points, facular points, bright points). 
The ``bright'' is because they appear bright in many diagnostics (more
below). 
They are not limited to active regions as spots and pores usually are. 
In these they do appear in denser concentrations but elsewhere they
are also present, including very quiet areas. 
There are fewer in the polar caps, but the ones there are important
gauges of polar fields.

Near the limb the small MCs are known as faculae. 
These were observed well before their small-bright-point signature
closer to disk center was recognized because they appear brighter in
slanted viewing.
Faculae were already visible towards the limb in white light images
that did not reach the high resolution of modern telescopes. 
This fortunate visibility enables using the number of polar faculae as
polar field proxy, using Sheeley's inventories of their numbers on
historical Mount Wilson full-disk photographs.
 
In principle the best way to observe MCs is through their strong-field
signature in Zeeman line splitting or broadening. 
However, this fails towards the limb where the radial MC fields become
less aligned with the line of sight as required for
circular-polarization Zeeman signature.

However, also on the disk magnetograms sampling Zeeman diagnostics
generally show them incompletely. 
The Helioseismic and Magnetic Imager (HMI) onboard the Solar Dynamics
Observatory (SDO) collects full-disk magnetograms every 45 seconds
since the spring of 2010, but while these monitor the distribution of
the stronger concentrations over the visible disk continuously, their
angular resolution and magnetic sensitivity are low so that they
detect only larger ones and do not resolve these.

The mid-ultraviolet images (1600\,\AA\ and 1700\,\AA) of the
Atmospheric Imaging Assembly (AIA) onboard SDO chart these small
magnetic concentrations more completely in the form of bright grains,
but without polarity distinction as magnetograms do. 
The resolution is as bad as for HMI but the MC inventory is more
complete through higher contrast, especially towards the solar limb.

Figure~\ref{fig:rrfig1} shows simultaneous images from HMI and AIA at high,
medium, and low levels of activity. 
Each image contains magnified cut-outs to show detail. 
Only the top continuum image in the first column contains sizable
spots, primarily in the southern activity belt. 
Elsewhere it shows granulation. 
The corresponding magnetogram in the second column show extended
groupings of small black or white (opposite polarity) grains marking
MCs. 
They are densest where there is most activity and there display the
Hale polarity laws. The dense groups around spots constitute
active-region plage, but away from spots there are also extended
clusters with a preferred polarity, also called plage or ``active
network''. 
They display tilts according to Joy's law. 
Similar ones appear in the second magnetogram but in the bottom
magnetogram there are no extended groupings. 
The insets show that the plages have roughly cellular ordering
constituting magnetic network. 
The top-right inset in the second magnetogram shows this network as
mostly unipolar (white) but with a few opposite-polarity (black)
``internetwork'' grains in the cell interiors.

Towards the poles the magnetograms show less network patterning and
more bipolar salt-and-pepper MC sprinkling, but still with some
preference for small-scale polarity sharing.

The ultraviolet 1700\,\AA\ images from AIA in the third column
illustrate that all magnetic elements appear as bright grains at this
wavelength. 
The weaker 1700\,\AA\ grains between these mark acoustic waves fed by
the solar surface oscillations. 
Comparing the various insets with the corresponding magnetogram insets
shows nearly 1:1 pattern equality (apart from the polarity sign)
between bright ultraviolet grains and kilogauss MCs.

\paragraphrr{Formation and brightness of small magnetic concentrations}

\begin{figure*}[!hbtp]
  \centering
  \includegraphics[width=17cm]{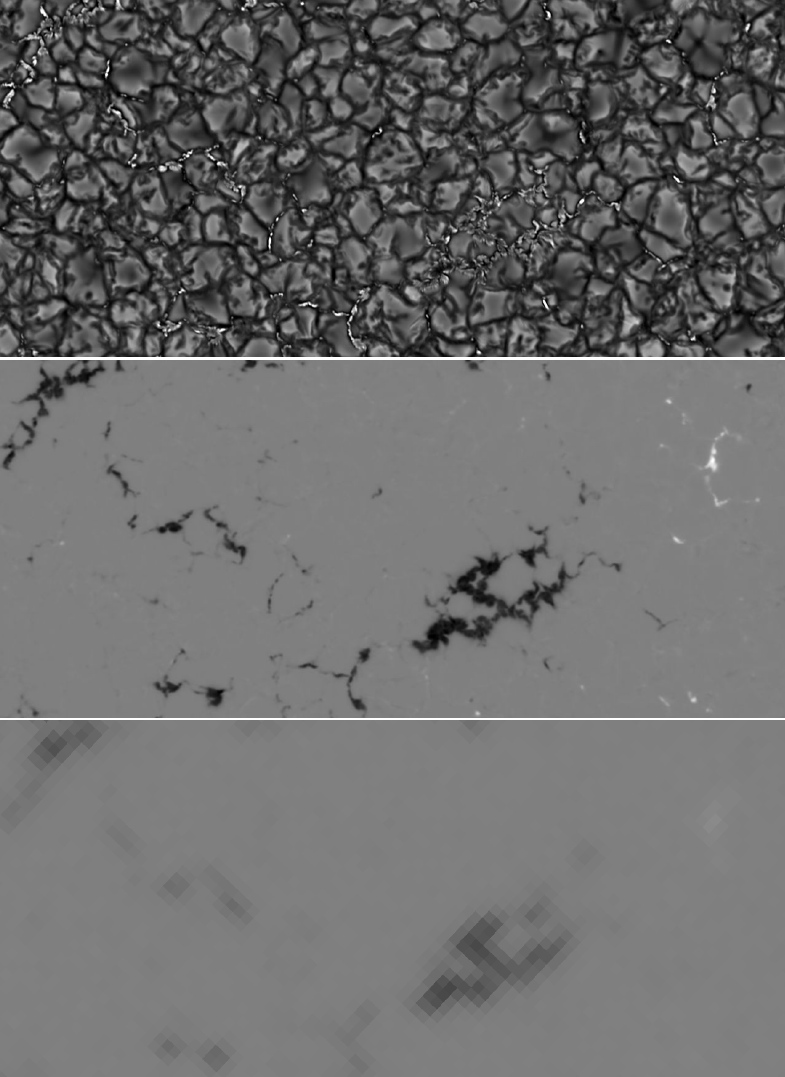}
  \caption[]{\label{fig:rrfig2} %
  Granulation and magnetic concentrations in a small quiet-Sun area
  near disk center. 
  Image sizes: $48\times22$\,arcsec or $35,400\times16,100$\,km. 
  The upper two panels are cut-outs of larger images taken by
  V.M.J.~Henriques with the Swedish 1-m Solar Telescope (SST) and
  described and analyzed by \citetads{2019A&A...631L...5B}. 
  Top panel: intensity at 3950\,\AA\ with bandwith 13.2\,\AA. 
  Middle panel: corresponding map of the line-of-sight magnetic field
  derived with an inversion technique from simultaneous polarimetry in
  the \FeI\,6301.5 and 6302.5\,\AA\ lines. 
  Bottom panel: corresponding cut-out from a simultaneous SDO/HMI
  full-disk magnetogram. 
  Its pixels are 16 times larger linearly. 
  The greyscale of both magnetograms is clipped at 1500~Gauss absolute
  strength. 
  SST data courtesy of L.H.M. Rouppe van der Voort.
  }
\end{figure*}

Much higher resolution than in Fig.\,\ref{fig:rrfig1} is obtained at
the best ground-based telescopes, in images in which these small
magnetic concentrations appear bright with respect to their
surroundings. 
Most reside in intergranular lanes so they become invisible at low
resolution because their local brightness cancels against the
surrounding darkness. 
Figure~\ref{fig:rrfig2} shows the state of the art with observations from the
Swedish 1-m Solar Telescope (SST), currently the best ground-based
solar telescope by combining an outstanding site, superb optics,
vacuum technology, advanced adaptive wavefront correction, further
numerical image restoration, and fast high-quality Fabry-P\'erot imaging
spectrometers operating at red and at violet wavelengths
(\citeads{2019A&A...626A..55S}).  

The top panel shows surface granulation. 
Within some intergranular lanes there are numerous small bright
features, clearly not ``points'' but with varying shape, often
elongated, that generally follow the morphology of the intergranular
lane in which they reside (and follow with time in image sequences). 
The SST magnetic map in the second panel shows that this small scene
contains mostly negative-polarity network, including a somewhat larger
``enhanced network'' patch with larger bright point density. 
There is some positive-polarity network at right. 
There are very few magnetic concentrations above the sensitivity
threshold in the internetwork areas; also these have corresponding
bright features in the top panel.
  
The two magnetograms are both clipped at 1500~Gauss absolute field
strength. 
The magnetic features appear much weaker in the HMI magnetogram
because they are much smaller than the pixels over which their
averaged magnetic signature is measured, as demonstrated by the
smallness of the bright points in the top panel. 
This comparison shows the poor rendering of small-scale magnetism in
HMI magnetograms.
Only the larger field patches remain diffusely visible. 
Thus, the magnetograms in Fig.\,\ref{fig:rrfig1} show tips of MC
icebergs only. 
These were clipped at only 100~Gauss in order to display their
location and polarity more completely, not their actual field
strength. 
The SST field map comes closer to resolving them, with many pixels
showing apparent field strengths above 1000~Gauss.

The nature of these small MCs and why they appear bright are well
understood. 
Just as in umbrae the strong magnetic field suppresses turbulent
convection locally, so that outward convective energy transport from
below is inhibited which makes them cooler than their surroundings. 
The strong field also contributes high magnetic pressure. 
Persistent presence of such strong-field features (they often last
hours) requires correspondingly low gas pressure inside so that the
sum of both pressures balances the outside field-free gas pressure
(``magnetostatic equilibrium'',
\citeads{1976SoPh...50..269S}). 

In contrast to sunspot umbrae and pores, the small MCs do not appear
darker but brighter than their surroundings. 
Naturally this was initially attributed to unidentified heating, but
from their magnetostatic modeling
\citetads{1976SoPh...50..269S} 
and \citetads{1981SoPh...70..207S} 
suggested that the apparent brightening comes from the hot walls of
the tube-like MCs below the surface. 
Their insides are relatively cool from the magnetic suppression of
outward convective energy flow, but the hot-wall radiation scatters
out to produce a bright feature if the MC is small enough, whereas
this contribution becomes negligible for larger pores and umbrae. 
Towards the limb the viewing along the slanted line of sight passes
further through the relatively empty fluxtube than besides that, and
so samples the hot granule behind causing the enhanced contrast of
limbward faculae with stalk-like appearance.

Subsequently, the kilogauss strength of MC fields was established with
dual-line spectropolarimetry by 
\citetads{1978A&A....70..789F}, 
verified in a number of studies (reviewed by
\citeads{1993SSRv...63....1S}), 
and then followed by detailed
numerical MHD simulations 
(e.g., \citeads{2004ApJ...607L..59K}; 
\citeads{2004ApJ...610L.137C}) 
in which magnetic concentrations
appear and brighten very similar to those in the best observations,
also as limb faculae in slanted viewing. 
These simulations insert uniform magnetic field of a few hundred
Gauss, horizontal or vertical, throughout or at the bottom of a
well-developed purely hydrodynamic simulation and then follow its
reconfiguration and shredding by the continuing turbulent convection. 
Below the surface the gas pressure generally dominates over the
magnetic pressure so that the fields are forced to follow the gas
motions (in the low-density corona this is reversed so that gas in
coronal loops is frozen-in to the field). 
The uniform initial field is so quickly transformed into more or less
vertical threads that are expelled by the convective flows from
granules to their surrounding lanes, and then swept to mesogranular
boundaries.

The simulations are yet too small in volume to harbor larger
supergranulation, but the observed network patterning which outlines
boundaries between supergranulation flow cells implies that subsequent
sweeping to and collection in supergranular boundaries follows (more
below). 
In summary, the simulations show that any field moving from somewhere
into the turbulent convection just below the surface is transformed
into kilogauss concentrations that are swept along by the local
granular, mesogranular and supergranular flows.

The low internal gas pressure enhances apparent MC brightness in
spectral diagnostics that are density-sensitive. 
Atomic lines such as the optical ones from Fe I weaken or even vanish
in MCs because most or all iron becomes ionized within them; in the
older literature this small-scale vanishing was called ``line gaps''. 
Molecular bands as the Fraunhofer-named G-band of CH molecules near
4310\,\AA\ and the CN band below 3884\,\AA\ weaken from enhanced
dissociation. 
The first convincing bright-point MC observations were therefore done
(at Pic du Midi) in the G band which permits imaging with relative
wide (10\,\AA) spectral bandpass, hence high signal-to-noise. 
In the outer damping wings of the Balmer lines and the strong \CaII\
lines the opacity diminishes from smaller density-sensitive
collisional broadening; this is the case in the top panel of
Fig.\,\ref{fig:rrfig2} which samples wavelengths in the overlapping outer
damping wings of \CaIIHK\ between these lines. 
In the mid-ultraviolet continua sampled by AIA the MCs show enhanced
brightening from ionization of \FeI\  and \MgI\ which contribute most
continuous opacity at these wavelengths; this brightening is evident
in the third column of Fig.\,\ref{fig:rrfig1}.
For all these diagnostics the smaller opacity within the MCs implies
deeper apparent holes for outward hot-wall radiation, hence a larger
contribution of that, and also larger transparency in facular viewing
along slanted lines of sight close to the limb that so penetrate
further into the hotter granules behind the MCs and sample more of
their brightness.

\paragraphrr{Weak internetwork fields}

The kilogauss MCs that constitute network and plage are smaller than
the spots and pores in active regions, but they do not represent the
smallest or weakest in the hierarchy of solar magnetic surface
features. 
In recent years it has become clear that also within quiet-Sun
supergranular cells weak ``internetwork'' fields are copiously present
in very tangled and dynamic form on granular scales. 
The first telltale was measurement of Hanle depolarization
(\citeads{2004Natur.430..326T}), 
followed by detection of abundant primarily horizontal fields at
granular scales in full-Stokes spectropolarimetry with the Hinode
satellite (Lites et al.\ 
\citeyearads{2008ApJ...672.1237L}, 
\citeyearads{2017ApJ...835...14L}) 
of which the 50-cm aperture furnishes the highest resolution and
sensitivity in magnetogram sequences from space so far. 
The absence of fast-varying atmospheric image distortion (``seeing'')
in space permits much longer integration times than for groundbased
telescopes, resulting in much higher sensitivity and hence weaker
field detection than possible with SST and HMI magnetometry as in
Fig.\,\ref{fig:rrfig2}.

Numerical MHD simulations (see
\citeads{2010ApJ...714.1606P}) 
have established that such weak small-scale quiet-Sun fields likely
result mostly from local convection near the surface operating as a
small-scale dynamo, less from convective shredding of stronger
kilogauss network/plage MCs or from shredding yet larger preceding
active regions. 
These weak internetwork fields are relatively stronger and more
vertical in intergranular lanes. 
Higher up, but still within the low photosphere, they close across
granules in tiny horizontal loops. 
The latter loops cover more area than the lane fields so that the
dominating weak-field signature in Hinode data is horizontal. 
This small-scale near-surface dynamo action is inherent in the
turbulent convection producing the granulation and does not vary with
latitude or the activity cycle.

\begin{figure*}[hbtp]
  \centering
  \includegraphics[width=\textwidth]{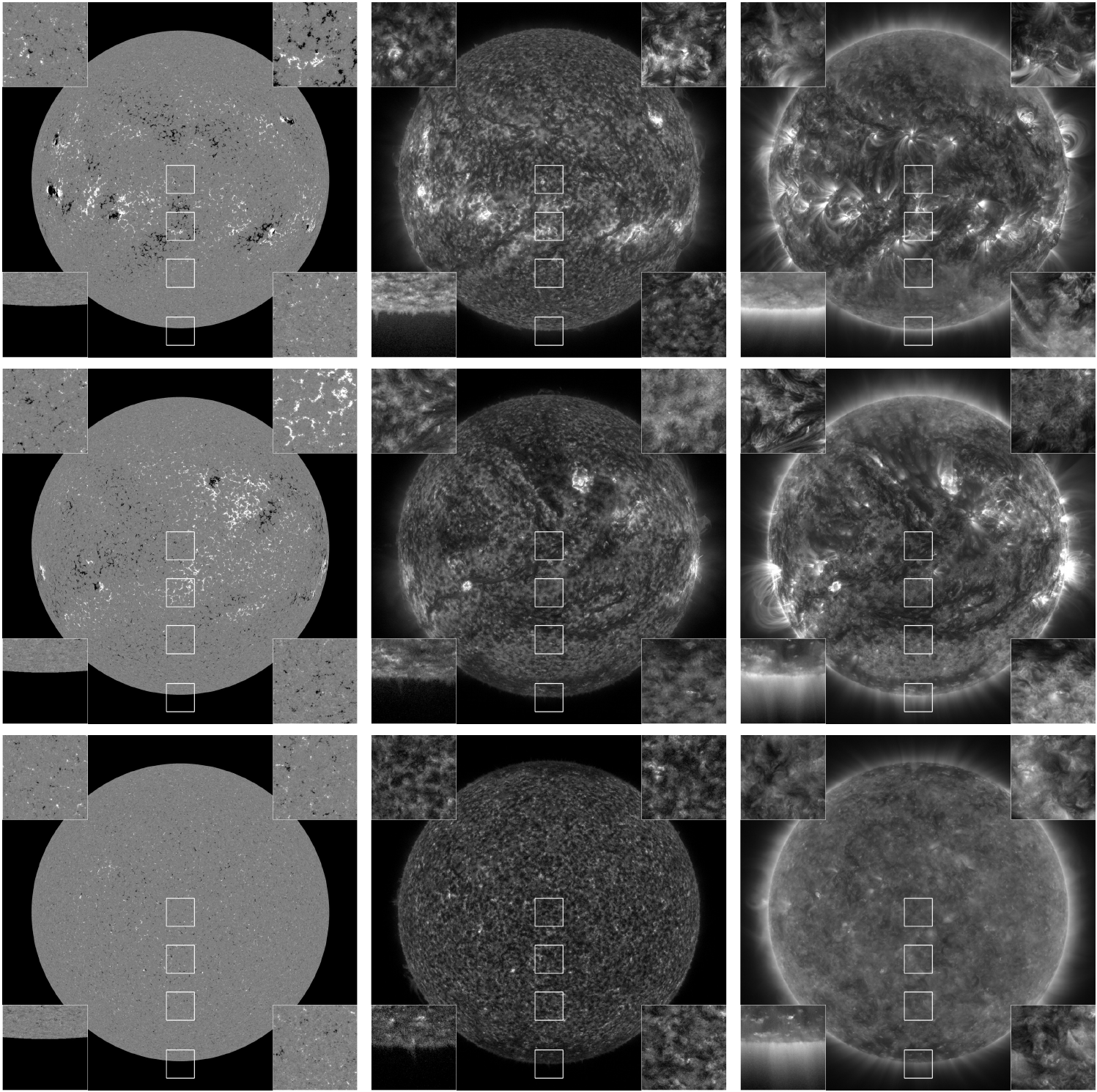}
  \caption[]{\label{fig:rrfig3} %
  Companion to Fig.\,\ref{fig:rrfig1} showing the corresponding
  chromosphere and corona for the same observing moments sampled in
  the three rows.
  The first column repeats the HMI magnetograms in the center column
  of Fig.\,\ref{fig:rrfig1}. 
  The second column shows the overlying chromosphere with AIA
  304\.\AA\ images which sample the \HeII\ Lyman-alpha line formed at
  temperatures around 100,000\,K. 
  The third column shows the overlying corona with AIA 171\,\AA\
  images which sample \FeIX\ and \FeX\ lines formed at temperatures
  around one million Kelvin. 
  The brightest parts of the AIA images are clipped to enhance quieter
  areas. 
  The full-disk images are less cropped than in Fig.\,\ref{fig:rrfig1}
  to admit off-limb features in the 171\,\AA\ images. 
  The fourth cut-out (at bottom left per panel) therefore includes the
  limb.
  }
\end{figure*}

\paragraphrr{Network and plage patterns}

Small kilogauss concentrations and weak internetwork fields are the
producers of larger-scale network and plage patterns on the solar
surface as seen in the magnetograms of Fig.\,\ref{fig:rrfig1}. 
Such production was first established for the kilogauss concentrations
called ephemeral regions 
(\citeads{1973SoPh...32..389H}). 
They are small bipolar kilogauss MC pairs that emerge from deeper
layers, presumably as tiny Omega loops in the overall toroidal field
system that rise buoyantly through the surface, similarly to the much
larger ones that produce active regions with bipolar spot pairs. 
They seem to emerge randomly in the activity belts but not near the
poles. 
The emerging pairs are most easily detected in internetwork regions
where they pop up in isolation. 
They are then seen to split; the two components are then transported
in random-walk fashion along with the supergranular flows to the cell
boundaries. 
Underway they frequently suffer same-polarity merging and
opposite-polarity cancelation, but enough survive and make it to the
network to replenish all its magnetic flux on a time scale of one to a
few days, while a similar amount of flux vanishes though cancelation
(\citeads{1997ApJ...487..424S}). 
 
Merging enhances their local field strength. 
Apparent cancelation can be due to fields with U-loop shape rising
through and out of the surface, with the observed surface
cross-sections of their legs appearing to move together and vanishing. 
Similarly, apparent cancelation may result from fields with inverted
U-loop shape sinking below the surface. 
Other cancelation can occur in small-scale magnetic reconnection in
which only the net surplus of the polarity with larger flux remains
visible. 
Operating at small scales, these processes are probably also important
in shredding and converting larger-scale magnetic fields to the small
scales where Ohmic dissipation can remove them, hence in doing away
with the flux that surfaces at large activity and is not spilled into
space by CMEs etc. 
A good discussion of these processes and their role in flux spreading
and removal is given in Chapt.\,6 of 
\citetads{2000ssma.book.....S}.  

Numerical simulations suggest that the weak granular internetwork
fields found with Hinode are insufficient to play a direct role in the
overlying chromosphere 
(\citeads{2019ApJ...878...40M}). 
However, 
\citetads{2014ApJ...797...49G} 
have used sensitive multi-day Hinode
spectropolarimetry to show that also these weak features tend to drift
per random walk to the supergranular cell boundaries, suffering much
merging and cancelation underway. 
In merging the combined weak field may become strong enough to
suppress internal convection and ``collapse'' into kilogauss tube-like
MCs which may so contribute to strong-field network as seen in
Fig.\,\ref{fig:rrfig1}. 
\citetads{2014ApJ...797...49G} 
found that weak-field merging dominates markedly
over cancelation and that the resulting net quiet-field contribution
can replenish all network flux within one day, competitive with
ephemeral regions. 
The picture so emerges that larger-scale unipolar network and plage
patterns are found where equatorial field in the form of ephemeral
regions or polar field in the form of polar faculae win the
competition with the weak-field granular dynamo producing
salt-and-pepper bipolar network.

The full-disk magnetograms and their cut-out enlargements in the
second column of Fig.\,\ref{fig:rrfig1} demonstrate resulting strong-field
plage and network patterns. 
The high activity sampled in the first-row magnetogram has extended
dominating-polarity plage areas along the activity belts but fewer MCs
with more salt-and-pepper appearance and less cellular patterning in
faculae towards the poles.
The medium activity sampled in the second magnetogram shows similar
but lower amplitude patterns. 
The low activity sampled in the bottom magnetogram shows
salt-and-pepper sprinkling all over the disk, with only one minor
active region in the North-East.

Quiet bipolar network produced by local dynamo action on granular
scales does not vary with cycle phase or latitude. 
Ephemeral regions tend to emerge preferentially in the activity belts
and produce bipolar network in addition to the more active unipolar
network and plage found in and near active regions that resulted from
much larger toroidal-system loop emergence. 
Thus, active region network and plage as well as quieter network
produced by ephemeral regions occupies the activity belts and is
modulated along with the cycle, with the sunspot number a valid proxy. 
Similarly, where polar field emergence dominates over the granular
dynamo the number of polar faculae represents a valid proxy
(\citeads{2012ApJ...753..146M}). 

\paragraphrr{Small-scale magnetism and the heliosphere}

Figure \ref{fig:rrfig3} is a companion to Fig.\,\ref{fig:rrfig1}. 
It shows corresponding heating patterns in the outer solar atmosphere:
the chromosphere in the middle column, the corona in the third column.
For these diagnostics brightness implies the presence of 100\,000 K
gas (304\,\AA) and of 1 million K gas (171\,\AA), respectively.

The top row sampling high activity shows activity belts dominated by
active regions, especially on the Southern hemisphere. 
These harbor dense clusters of 304\,\AA\ emission (here truncated by
greyscale clipping). 
In 171\,\AA\ they show much closed connectivity in the form of coronal
loops. 
Towards the poles the heating appears more homogeneous, showing grainy
patterns in 304\,\AA\ that correspond fairly close to the underlying
magnetic network patterns (compare the 304\,\AA insets with the
magnetogram insets). 
Towards the poles the 171\,\AA image shows more diffuse million-K
brightness with many polar plumes (bright outward stalks). 
The 171\,\AA\ cut-outs retain only slight and diffuse underlying
network signature.

The middle row shows similar scenes but with an extra component:
coronal holes that are visible in both the chromospheric and coronal
images. 
They include an extended polar cap one at the South pole.
Comparison with the magnetogram at left shows that the holes tend to
neighbor unipolar plage. 
The 171\,\AA\ insets show better (but diffuse) correspondence with the
304\,\AA\ ones here.

The bottom low-activity row shows more common regular polar-cap holes,
in the chromosphere regular network-size blobs everywhere, roughly
representing a map of the supergranulation. 
The corona shows much less loop connectivity but yet there is
1-million K heating everywhere, with isolated grainy patches of extra
heating of which some coincide with bright 304\,\AA\ enhanced-heating
grains.

The precise mechanisms through which small-scale MCs govern this
overlying heating are not yet identified, but most likely they are a
mixture of magnetic reconnection and Alvenic wave and shock generation
and dissipation. 
In larger-scale eruptions as flares and surges reconnection plays a
key role. 
On the smaller scales discussed here it is now well established that
small Ellerman bombs in active regions are due to opposite-polarity MC
cancelation marking strong-field reconnection, both observationally
(\citeads{2011ApJ...736...71W} 
and with numerical simulations 
(\citeads{2019A&A...626A..33H}). 
In quiet network well away from active regions similar but smaller
``Ellerman-bomb like'' reconnective cancelation events were detected
with the SST 
(\citeads{2016A&A...592A.100R}) 
and then also in
simulations 
(\citeads{2017A&A...601A.122D}). 

Much more ubiquitous are so-called ``spicules type II'' that emanate
from all network including monopolar network and also in coronal
holes; the latter are easier detected as off-limb spicules through
less closed-field confusion. 
Their tips may reach coronal temperatures (off-limb:
\citeads{2011Sci...331...55D}, 
on-disk: \citeads{2016ApJ...820..124H}). 
Recent simulations including ion-neutral separation
(\citeads{2018ApJ...860..116M}) 
suggest that these are not produced by kilogauss MC cancelation but as
shocks from tension release of complex tangled weaker fields. 
These spicules are probably a major contributor to quiet-Sun heating.

Within coronal holes, in particular polar ones, the production of
solar plumes that may also govern the fast solar wind is attributed to
opposite-polarity cancelation against network MCs 
(\citeads{1995ApJ...452..457W}). 

The largest-scale eruptive effect of network and plage outside active
regions concerns quiet-Sun filament formation and eruption into
coronal mass ejections (CME). 
Away from activity, filaments form above polarity dividing lines
between extended opposite-polarity regions as those in the
magnetograms in the top and center rows of Fig.\,\ref{fig:rrfig1}), in
particular dividers where closed fields arch away on both sides. 
Filaments may then live for months but suddenly snap into
CME-producing eruptions from tether cutting by pattern changes
including new flux emergence. 
The CME catalogs compiled and compared in detail by 
\citetads{2019SSRv..215...39L} 
show that from cycle minimum to maximum the overall CME frequency
increases tenfold, while the latitudes where they occur spread from
only near the equator to all latitudes, with polar crown filaments
during cycle maximum. 
Twice more CMEs result from erupting filaments than from flares, but
the highest source-region correlation is with coronal streamers. 
Thus, the large-scale polarity patterns as in the high-activity
magnetograms in Fig.\,\ref{fig:rrfig1} play a pivotal role. 
Such patterns are absent in the minimum-activity magnetogram at the
bottom. 
Indeed, the observed daily CME production rate tracks the sunspot
number quite well and without delay, so that both active-region and
quiet-Sun CME production can be handled with this proxy.

\bibliographystyle{aa-note} 
\interlinepenalty=10000
\bibliography{rjrfiles,adsfiles} 

\end{document}